\documentclass[12pt]{article}

\usepackage{amsmath,amsfonts,amssymb,bm}

\usepackage[utf8]{inputenc}
\usepackage{amsmath,amssymb}
\usepackage{latexsym}
\usepackage{slashed}
\usepackage{authblk}
\usepackage{xcolor}
\usepackage{hyperref}

\usepackage{cancel}
\usepackage{indentfirst}
\usepackage{graphicx}
\usepackage{epstopdf}
\usepackage{dcolumn}
\usepackage{bm}
\usepackage{lipsum}
\usepackage{slashed}
\usepackage{enumitem}
\usepackage{txfonts}
\usepackage{esint}
\usepackage{xcolor}
\usepackage{mathtools}
\usepackage{transparent}
\usepackage{upgreek}
\usepackage{tabularx}
\usepackage{csquotes}

\begin{document}
\title{Pion in the Bethe-Salpeter approach with separable kernel}

\author{ 
	\parbox{\linewidth}{\centering
          S. G.~Bondarenko$^{1,2}$,
          M. K.~Slautin$^{1,2}$\thanks{E-mail:~slautin@jinr.ru}
}
}

\date{}

\maketitle


\begin{center}
{
$^1$ \it Bogoliubov Laboratory of Theoretical Physics, JINR, Dubna, 141980 Russia\\
$^2$ \it Dubna State University, Dubna, 141980 Russia
}
\end{center}
 

\maketitle

\begin{abstract}
In the paper, the static and dynamic properties of the pion in the Bethe-Salpeter approach are considered.
The rank-one separable kernel of the quark-antiquark interaction is used to solve
the equation analytically. Multidimensional integrals describing  pion properties are calculated
by several numerical methods and compared with previous papers. An error in the calculation
of the interaction part of the elastic pion form factor is found. Using the corrected results, the kernel parameters are refitted. The calculated static properties as well as transition and 
elastic form factors are presented in comparison with the recent experimental data.
\end{abstract}

\noindent
PACS: 13.20.Cz, 13.25.Cq, 14.40.Aq

\section*{Introduction}

The pion is the simplest quark-antiquark system. The small mass of the pion, compared to the masses
of other mesons, allows the pion to play an important role in the description of nuclear dynamics.
There are many models for pion description:
sum rules in QCD with a hard scattering estimate~\cite{Nesterenko:1982gc}; nonrelativistic potential model~\cite{Godfrey:1985xj};
relativistic model using the light-front formalism~\cite{Jacob:1988as};
Nambu-Jona-Losinio model~\cite{Nambu:1961tp,Zhang:2024dhs,Anikin:1995cf,Bernard:1986ti,Hatsuda:1985ey};
model cased of chiral symmetry~\cite{Gross:1991te}; instanton pion model~\cite{Anikin:2000rq}; lattice calculations~\cite{ExtendedTwistedMass:2023hin},~\cite{Gerardin:2019vio};
recent models based on the Bethe-Salpeter equation
with dressed quark and gluon propagators~\cite{Maris:2000sk,Kekez:2020vfh,Hernandez-Pinto:2023yin}.

The study of the pion properties has recently become relevant due to the transition form factor of the pion that gives a contribution to the anomalous magnetic
moment of the muon (g-2)~\cite{Hernandez-Pinto:2023yin}. In addition, a number of experiments
are planned to measure the charge and transition form factor of the pion:
E12-22-003~\cite{E12-22-003}, E12-06-101~\cite{E12-06-101}, E12-19-006~\cite{E12-19-006},
PR12-16-003~\cite{PR12-16-003}.

In this paper, a model based on the relativistic covariant Bethe-Salpeter equation with
a separable kernel~\cite{Ito:1991pv,Ito:1991sz} is considered. 
The choice of this model is due to the simplicity and analytical solution to the
pion vertex function.
The parameter $\Lambda$ characterizing the quark-antiquark interaction region manifests itself
in the separable interaction form factor and provides a close connection between the pion composite
structure and the quark mass.

The paper is organized as follows: in Section 1 the basic formulae 
of the formalism are given, in Section 2 the pion constants and form factors
are described, in Section 3 the model parameters and the obtained results are
discussed, and in Section 4 the main results are summarized.

\section{The solution for a pion}

The Bethe-Salpeter equation for the pion vertex function is written as follows:
\begin{equation}
\Gamma_{\alpha\beta}(k;p)=i\int\dfrac{d^4k''}{(2\pi)^4}V_{\alpha\beta:\epsilon\lambda}(k,k'';p)S_{\lambda\eta}(k''+p/2)\Gamma_{\eta\zeta}(k'';p)S_{\zeta\epsilon}(k''-p/2).
\label{eqgamma}
\end{equation}
where $p,k$ 
are the total and relative 4-momenta, respectively, and $V(k',k;p)$ is the interaction kernel. 
The pion mass $m_{\pi}$ is fixed and
can be considered as an implicit model parameter $p^2=m_\pi^2$ with $m_\pi=140$ MeV.
The fermion propagator with mass $m$ is defined as
$S(k)=(\cancel{k}-m+i\epsilon)^{-1}$, and the Dirac indices are denoted by Greek symbols.

The rank-one separable kernel of interaction can be written in the following form:
\begin{equation}
    V_{\alpha\beta:\delta\gamma}(k',k;p)=\sum_{i=1,4}\Delta^i_{\alpha\beta}(k';p)\overline{\Delta}^i_{\delta\gamma}(k;p),
\end{equation}
with
$$
\Delta^i_{\alpha\beta}(k';p)=f_i({k'}^2,k'\cdot{p})\Omega^i_{\alpha\beta},
\qquad
\overline{\Delta}^i_{\alpha\beta}(k;p)=f_i({k}^2,k\cdot{p})\overline{\Omega}^i_{\alpha\beta}.
$$
The matrices $\Omega^i$ and $\overline{\Omega}^i$ represent the spin structure,
and  $f_i({k'}^2,k'\cdot{p})$ and $f_i({k}^2,k\cdot{p})$ are the scalar functions of the 
scalar products of the $k', k, p$ 4-momenta.

In the general case, the pion vertex function can be written as
\begin{equation}
\Gamma(k;p)=\gamma^5\left[f_1(k^2;k\cdot{p})+p_{\mu}\gamma^{\mu}f_2(k^2;k\cdot{p})+k_{\mu}\gamma^{\mu}f_3(k^2;k\cdot{p})+\sigma_{\mu\nu}k^{\mu}p^{\nu}f_4(k^2;k\cdot{p})\right],
\label{eqgammatot}
\end{equation}
where the matrix $\gamma^5$ represents the pseudoscalar nature of the pion.
In the present paper, only the first Dirac structure and $k^2$ dependence of the scalar function
are taken into account for the sake of simplicity.
The solution in this case is:
\begin{equation}
    \Gamma(k)={N\gamma^5}{f(k^2)},
\end{equation}
where the normalization constant $N$ is fixed from the charge conservation equation.
The simple form of the BS vertex function is justified by the small value of the pion mass, but the full form~(\ref{eqgammatot}) must be investigated somewhere.
The proportionality of the BS vertex function
to the trial function of the interaction kernel explains the choice of the
equation form~(\ref{eqgamma}).

The scalar part of the vertex function is taken in the simple monopole form:
\begin{equation}
f(k^2)=\dfrac{1}{k^2-\Lambda^2+i\epsilon}.
\end{equation}
The parameter $\Lambda$ (about hundreds of MeV) defining the effective range of interaction.

\section{The pion constants and form factors}

In the paper, the following pion constants are analyzed:
the weak decay constant $f_{\pi}$,
the two-photon decay width $\Gamma_{\pi^0\xrightarrow[]{}\gamma\gamma}$,
the transition radius $r_{\pi\gamma}$, and
the charge radius $<r^2_{\pi}>$.

The form factor of the $\gamma^*\pi^0\xrightarrow[]{}\gamma$ transition
$F_{\pi\gamma}$ and the elastic charge form factor consisting of two parts, relativistic impulse approximation (RIA) and interaction current (int) --
$F_{\pi}(q^2)=F^{\textrm{RIA}}_{\pi}(q^2)+F^{\textrm{int}}_{\pi}(q^2)$,
are calculated as well.

All expressions for the considered observables could be found in~\cite{Ito:1991pv,Ito:1991sz}.
For all constants except the elastic form factor, full
agreement of the obtained results with the previous ones is found.

While calculating the interaction part $F^{\textrm{int}}$ of the elastic form
factor a difference in the results is found.
To ensure that the numerical methods are good for 4-dimensional integrals,
three alternative numerical methods are used. The first one is based on the
Cauchy theorem and  calculates the residues on the $k_0$ poles that are determined by infinitesimal additions in the upper complex half-plane. Then the 3-dimensional integral on 3-momenta $k$ $\cos\theta_k$,
$\phi_k$ was calculated.
The second one is based on the Feynman parametrization and allows one to
perform 4-momenta $k$ integration and then calculates the integral on the Feynman parameters.
The third method is based on the Wick rotation procedure and
allows one to calculate the 4-dimensional integral in Euclidean space
with additional three-dimensional integrals at the points of the $k_0$ poles
that appear in the first and third quadrants ($k_0=i k_4$): 

\begin{equation}
    4\pi{i}\int dk_0\int_{0}^{+\infty}k^2dk \int_{-1}^{+1} d\cos\theta_k\, [f(k_0,k,\cos\theta_k)]\nonumber
     \end{equation}
    \begin{equation}
    =-4\pi\int_{-\infty}^{+\infty}dk_4\int_{0}^{+\infty}k^2dk \int_{-1}^{+1} d\cos\theta_k\, [f(k_4,k,\cos\theta_k)]\nonumber
    \end{equation}
    \begin{equation}
    +2\pi{i}\sum_n \int_{k_{\textrm{min}}}^{k_{\textrm{max}}}k^2dk \int_{\cos\theta_{k{\textrm{min}}}}^{\cos\theta_{k{\textrm{max}}}} d\cos\theta_k ,
 \textrm{Res}_n[f(k_0^n,k,\cos\theta_k)].\nonumber
\end{equation}

The results of the three above-mentioned methods were thoroughly compared.
In Fig.~1
a comparison of the described numerical methods
for $F^{\textrm{RIA}}_{\pi}$ and $F^{\textrm{int}}$ is shown, respectively.
The solid line represents the Wick rotation method results
without an additional pole contribution (WRMwp), 
the dashed line is the Cauchy theorem method (CTM), 
the dot-dashed line is the Feynman parameters method (FPM),
and the doted line is the full Wick rotation method (WRM).

As it is seen in the figure,
the form factors calculated by three numerical methods
agree within the statistical error.
For the Wick rotation procedure, two curves are presented: the total result (dot-dashed line) and the result without taking account of
additional 3-dimensional integrals (solid line)
in the first and third quadrants. 
The difference is obvious.

Also, a comparison of the obtained results with those in~\cite{Ito:1991pv,Ito:1991sz}
is shown in Fig.~2 for three sets 
of model parameters. The doted, solid, and dashed lines are our calculated results for sets I,II,III represented in Table 1, respectively. The dash-dot-doted, short doted, and dash-doted lines are the calculated results that were obtained in ~\cite{Ito:1991pv,Ito:1991sz} with the same sets of model parameters. The RIA part of the elastic form factor fully coincides, while
the interaction part has the opposite sign and different behavior.

\begin{figure}[htb]
\label{figFria265403}
\begin{minipage}[htb]{0.5\linewidth}
{\includegraphics[width=\linewidth]{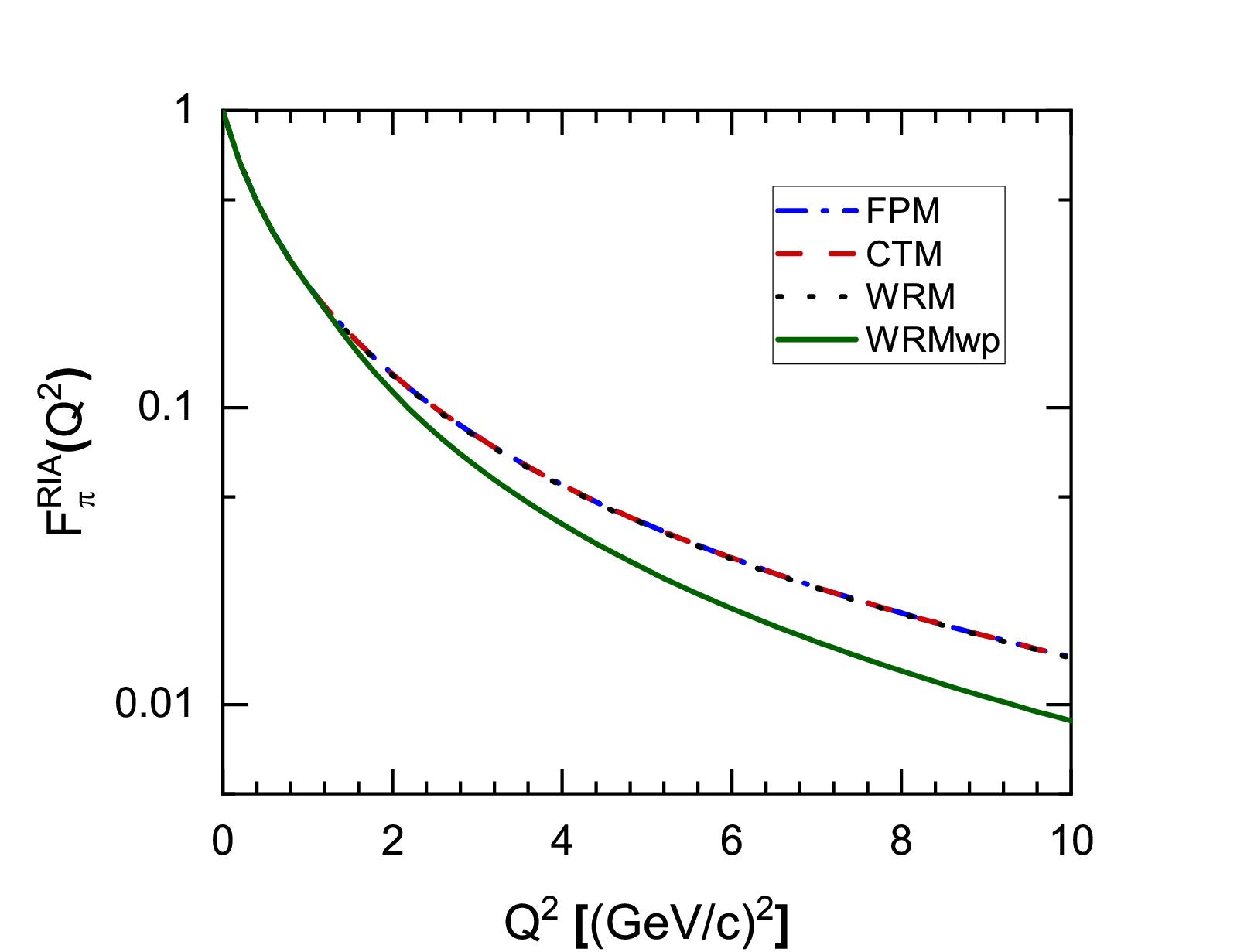}}
\end{minipage}
\begin{minipage}[htb]{0.5\linewidth}
{\includegraphics[width=\linewidth]{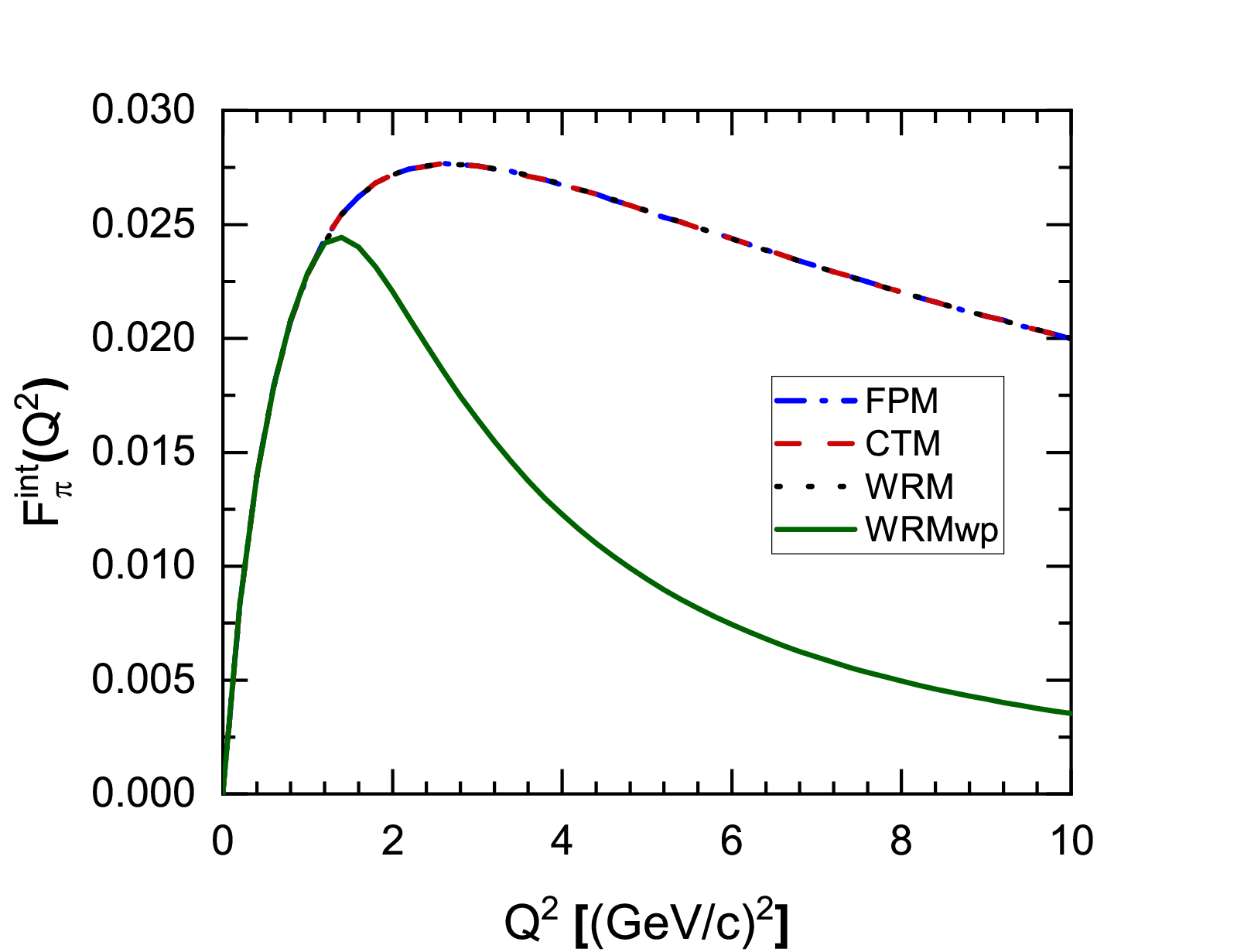}}
\end{minipage}
\caption{
A comparison of three numerical methods for the RIA (left panel) and interaction current (right panel) contributions to the charge pion form factor.}
\end{figure}

\begin{figure}[htb]
\begin{minipage}[htb]{0.5\linewidth}
{\includegraphics[width=\linewidth]{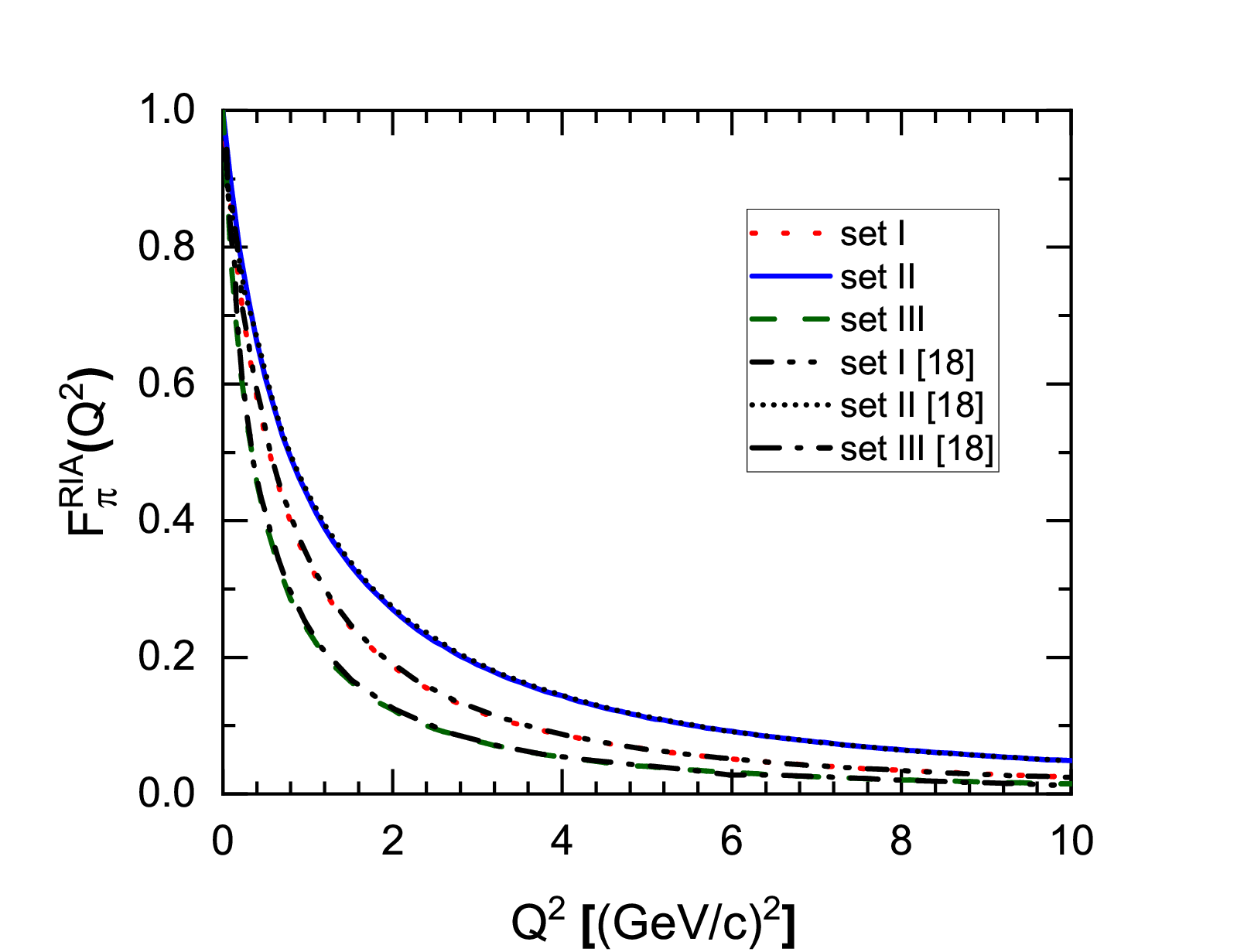}}
\end{minipage}
\begin{minipage}[htb]{0.5\linewidth}
{\includegraphics[width=\linewidth]{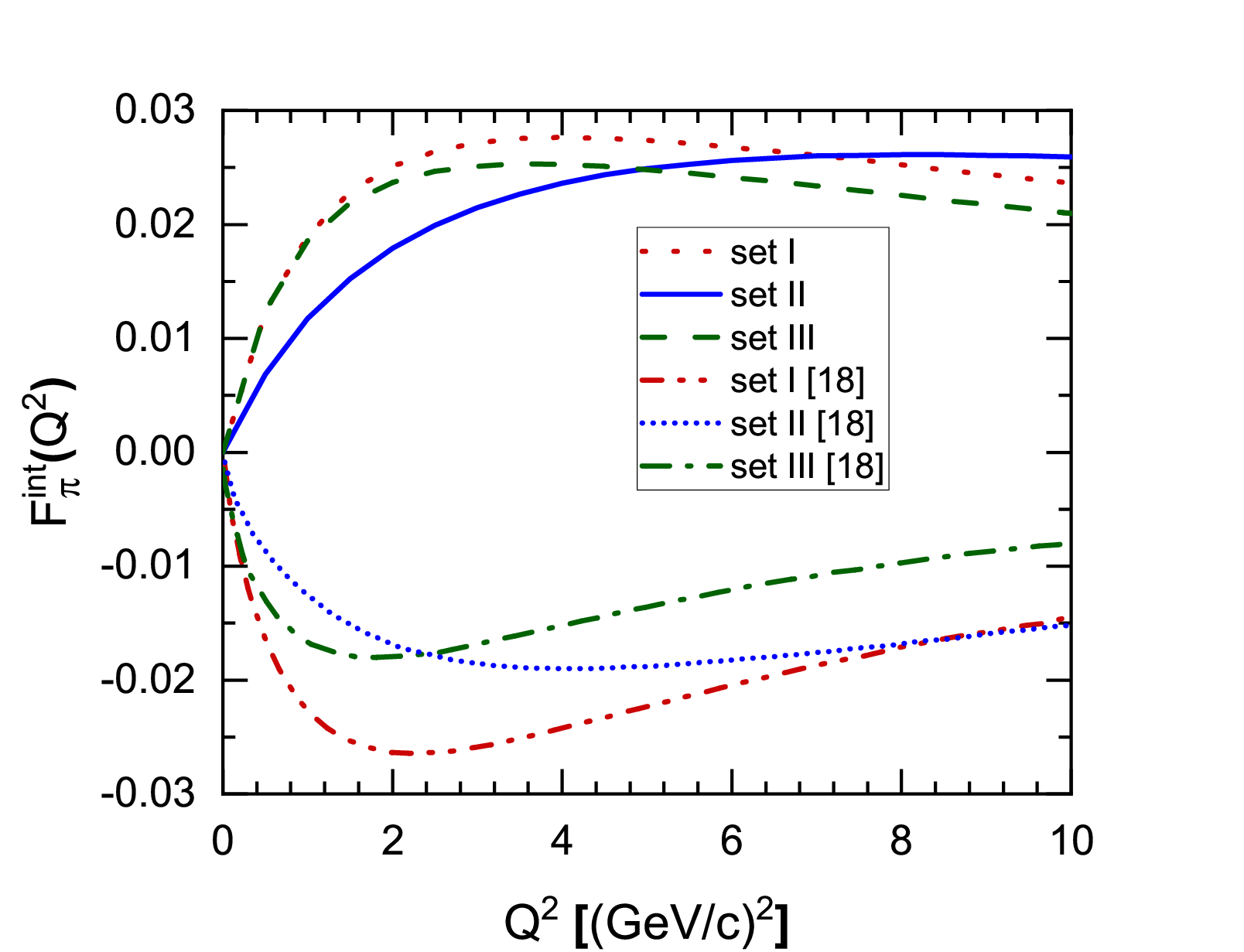}}
\end{minipage}
\caption{\label{figFria265403}
A comparison of the calculated results with those given in~\cite{Ito:1991pv}.
The RIA (left panel) and interaction current (right panel) contributions.}
\end{figure}

\section{Model parameters and results}

The model parameters (three sets) from~\cite{Ito:1991pv} have been used to calculate pion constants and compare them with the results of that paper.
Good agreement for all constants except the charge radius was found. 
The calculated values are shown in Table~\ref{tab1} -- sets I,II,III.
It is seen that the values for the constants differ a lot from the experimental data.
To reduce the difference, new sets of parameters were created.
The first one (set IV) was obtained by fitting only the $f_{\pi}$ and $r_{\pi\gamma}$ constants to describe the experimental data.
It perfectly describes two constants; however, it gives a bad value for $<r^2_{\pi}>$.
Therefore, a new set (V) was created by fitting the $f_{\pi}$, $r_{\pi\gamma}$,
$<r^2_{\pi}>$ and $\Gamma_{\pi^0\xrightarrow[]{}\gamma\gamma}$ constants.
It describes these constants with an accuracy of less than 10\% of the experimental ones.
The constituent quark mass $m$ is one of the two model parameters
which is fixed by pion static parameters.

\begin{table}[htb]
  \caption{Model parameters $m, \Lambda$ and observables
    $f_{\pi} , r_{\pi\gamma}, <r^2_{\pi}$>, $\Gamma_{\pi^0\xrightarrow[]{}\gamma\gamma}$}
    \label{tab1}
    \centering
    \begin{tabular}{ | l | l | l | l | l | l | l |}
 \hline
    & \small {$m$ (MeV)} & \small{$\Lambda$ (MeV)} & \small{$f_{\pi}$ (MeV)} & \small{$r_{\pi\gamma}$ (fm)} & \small{$<r^2_{\pi}>$ ($\textrm{fm}^2$)} & \small{$\Gamma_{\pi^0\xrightarrow[]{}\gamma\gamma}$ (eV)}  \\ \hline
    \small I & \small300.0 & \small500.0 & \small152.74 & \small0.575 & \small0.400 & 
    \small5.633\\ \hline
    \small II & \small300.0 & \small750.0 & \small175.00 & \small0.541 & \small0.308 & \small5.269  \\ \hline
 \small III & \small200.0 & \small500.0 & \small115.78 & \small0.815 & \small0.708 & \small 11.828 \\ \hline \hline 
  \small IV & \small265.0 & \small403.0 & \small130.46 & \small0.660 & \small0.549 & \small7.252 \\ \hline 
   \small V & \small260.0 & \small550.0 & \small143.37 & \small0.639 & \small0.459 & \small7.249  \\ \hline 
 \small exp &  &  & \small 130.41 $\pm$ 0.0002 & \small 0.659 $\pm$ 0.004 & \small 0.430 & \small 7.57 $\pm$ 0.03\\ \hline 
    \end{tabular}
\end{table}

The pion form factors were calculated using sets IV and V. Figure 3
shows the obtained results for the charge form factor $F_{\pi}(Q^2)$ (left panel:
the experimental data are from: 
triangle -- \cite{Bebek:1977pe}, circle -- \cite{JeffersonLabFpi:2000nlc},
rhombus -- \cite{JeffersonLabFpi-2:2006ysh}) and
for the transition form factor $F_{\pi\gamma}(Q^2)$ (right panel: the experimental 
data are from: rhombus -- \cite{CELLO:1990klc},  triangle -- \cite{CLEO:1997fho}, circle -- \cite{BaBar:2009rrj}, square --- \cite{Belle:2012wwz}) with experimental data.
The graph shows that both sets describe well the behavior of the transition form factor,
but set V better describes the latest data for the charge form factor.

\begin{figure}[htb]
\begin{minipage}[H]{0.5\linewidth}
  \center{\includegraphics[width=\linewidth]{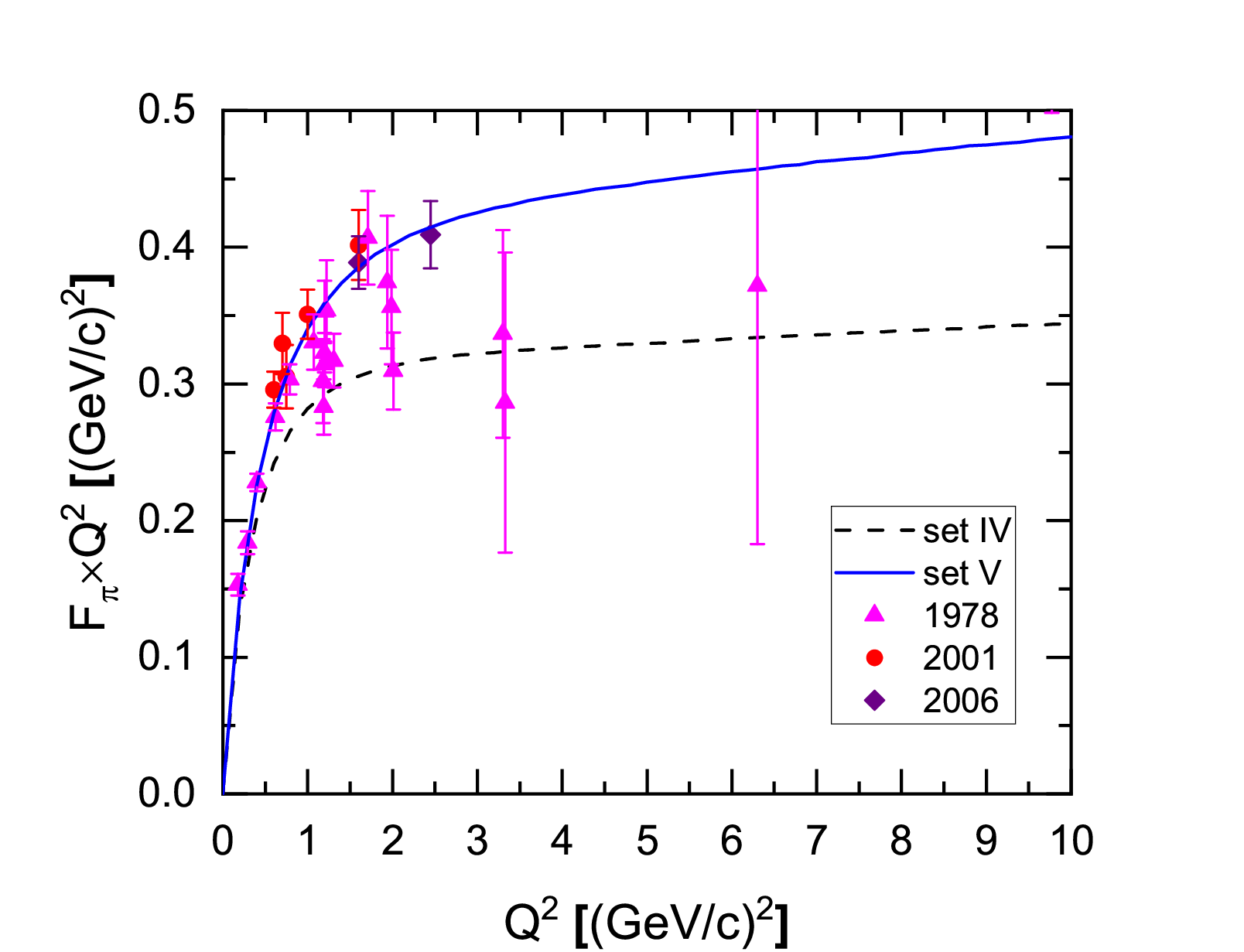}}
\end{minipage}
\begin{minipage}[H]{0.5\linewidth}
  \center{\includegraphics[width=\linewidth]{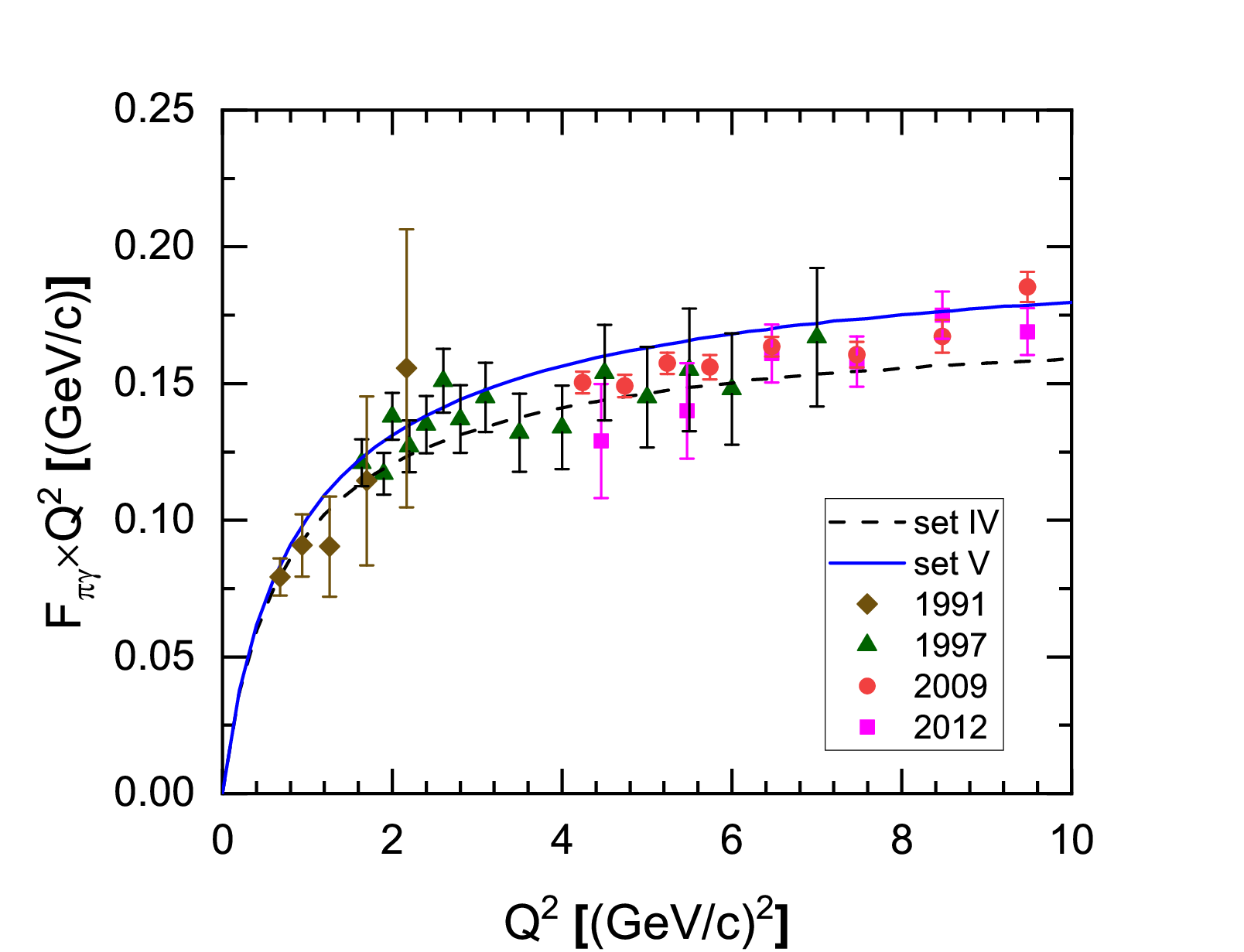}}
\end{minipage}
\caption{\label{FpiQ2exp4}
  Charge form factor $F_{\pi}\times{Q^2}$ (left panel) and
  transition form factor $F_{\pi\gamma}\times{Q^2}$ (right panel).}
\end{figure}

\section{Conclusion}

In this work, the model for the pion has been considered within the Bethe-Salpeter approach with a separable kernel of a quark-antiquark interaction.
The analytical solution for the pion vertex function was used 
to calculate the static and dynamic observables.

Three independent numerical methods were applied to calculate the elastic
pion form factor, and the difference with the previous results~\cite{Ito:1991pv} for the interaction current was found. Using the correct result for it, a new set of model parameters was obtained ($m = 260$ MeV, $\Lambda = 550$ MeV).
This set describes the static constants with a difference of less than 10\% from the experimental data. The pion form factors (elastic and transition) are in good agreement
with the latest experimental data.

The results are very sensitive to the values of the model parameters. It would also be good to investigate the influence of the form of the radial part of the vertex function.

\section*{FUNDING}
This work was supported by ongoing institutional funding. No additional grants to carry out or direct this particular research were obtained.

\section*{CONFLICT OF INTEREST}
The authors of this work declare that they have no conflicts of interest.

\bibliographystyle{pepan}

\end{document}